\documentclass[draft,showpacs,aps,prd,amsmath,amssymb, preprint]{revtex4}

\begin{document}

\title{Quark spin coupling in baryons - revisited}

\author{J. Chizma}
\author{G. Karl}
\address{Department of Physics, University of Guelph,
 Guelph, Ontario, Canada, N1G 2W1}

\date{\today}
\begin{abstract}
A direct connection can be made between mixing angles in negative
parity baryons and the spin coupling of constituent quarks. The
mixing angles do not depend on spectral data. These angles are
recalculated for gluon exchange and pion exchange between quarks.
For pion exchange the results of Glozman and Riska are corrected.
The experimental data on mixing are very similar to those derived
from gluon exchange but substantially different from the values
obtained for pion exchange.
\end{abstract}


\maketitle

\section{\label{intro}Introduction}
The spin-spin coupling between two fermions has two terms: a
``tensor'' term and a ``contact'' term. In atoms, electron spins
interact with the nuclear spin and this explains the hyperfine
structure: the tensor term is ordinary magnetic dipole-dipole
interaction, and the contact term is part of the same interaction
when the dipoles are at the same point. In nuclei, nucleon spins
interact through pion exchange and there is a similar coupling
with a different weight for the contact term. For constituent
quarks in baryons there is a controversy in the literature between
``gluon exchange'' (OGE) which mimics the magnetic coupling and
``pion exchange'' (OPE). In the early days of constituent quarks
OGE was applied to ground state \cite{georgi} and excited baryons
\cite{gromes, karl, isgur} with some success. Pion exchange was
also tried particularly in the context of bag models \cite{brown}.
More recently it has been argued that the entire spin dependent
coupling between constituent quarks is due to Goldstone Boson
Exchange \cite{glozman}, a generalization of OPE. This proposal
was criticized \cite{nathan, gloz}.

We deal here with a single issue: the mixing of states in the
lowest mass negative parity nucleons. These negative parity
nucleons have internal orbital angular momentum L=1, which couples
with an overall quark spin of S = 1/2 or 3/2 to give the total
angular momentum J. The physical states of J = 1/2 (or 3/2) are
mixtures of doublet and quartet spin states, and this mixing can
be determined from decay data. This issue has been discussed
already \cite{nathan, gloz}. However, we find that the discussion
has been flawed since \cite{nathan} used the estimates of
\cite{glozman} and the estimates of \cite{glozman} are based on
fitting the experimental mass spectrum. However, as we show below,
these mixing angles are independent of the mass spectrum and
depend only on the coupling and wavefunctions. We re-evaluate
these angles and find significant changes from those appearing in
\cite{glozman, nathan}. The differences between OPE and OGE become
larger and the data favors more clearly OGE, the same coupling as
between electrons and nuclei (ie., magnetic dipole type hyperfine
interactions). All this will be discussed in detail below, as well
as some items in the literature.

\section{\label{qqhi}Quark-Quark Hyperfine Interactions}

Our discussion here follows \cite{karl}, although for the sake of
clarity we repeat some of the material. \\
\subsection{\label{oge}One Gluon Exchange (OGE)}

This effective hyperfine interaction between two quarks in a
baryon has the form of magnetic dipole-dipole interaction in
Electrodynamics, (with dipoles produced by current loops, see
\cite{jackson}):
\begin{equation}\label{1}
H_{{\rm OGE}} = A \lbrace(8 \pi/3) \: \: \vec{S_{1}} \cdot
\vec{S_{2}}\: \:  \delta ^{3} (\vec{\rho}) + (3\vec{S_{1}} \cdot
\hat{\rho} \: \: \: \vec{S_{2}} \cdot \hat{\rho} - \vec{S_{1}}
\cdot \vec{S_{2}})\rho ^{-3} \rbrace.
\end{equation}
Here \(S_{1}, S_{2}\) are the spins of the two quarks, \(\sqrt{2}
\vec{\rho} = \vec{r_{1}} - \vec{r_{2}}\) is a vector joining them
and \(\hat{\rho} = \vec{\rho} / \vert \vec{\rho} \vert \) is a
unit vector and A is an overall constant which determines the
strength of the interaction. We do not need the value of A in what
follows, since we do not engage in fitting spectra (but A\(>0\)).
Nor does it matter if the value of A is too large to be
interpreted as single gluon exchange. The first term is called the
(Fermi) contact term and the second is the ``tensor'' term, but
these names obscure the origin of the second term which is the
ordinary dipole-dipole interaction for two separated dipoles of
spin one half. Recall that the contact term only contributes when
the two dipoles are in an orbital s-wave state (\(l_{12} = 0\)),
while the tensor term only contributes when the two dipoles are in
an orbital state with \(l_{12}\) different from zero ({\it unity}
here). It is also important to note that these two terms are parts
of the same physical interaction. In Eq.\ (\ref{1}) one assumes
that
the quarks are point-like. \\

\subsection{\label{ope}One Pion Exchange (OPE)}

Here we assume that the two quarks interact by exchanging a
massless pseudoscalar, the ``pion'', and the coupling takes the
form \cite{glozman}:
\begin{equation}\label{2}
H_{{\rm OPE}} = B \lbrace(-4 \pi/3) \: \: \vec{S_{1}} \cdot
\vec{S_{2}}\: \:  \delta ^{3} (\vec{\rho}) + (3\vec{S_{1}} \cdot
\hat{\rho} \: \: \: \vec{S_{2}} \cdot \hat{\rho} - \vec{S_{1}}
\cdot \vec{S_{2}})\rho ^{-3} \rbrace \vec{\lambda}_1 ^{f} \cdot
\vec{\lambda}_2 ^{f},
\end{equation}
where B is another constant and \(\vec{\lambda} _{1,2}^f \) are
the eight \(3 \times 3\) Gell-Mann SU(3) flavor matrices for
quarks number 1 and 2. If we consider strictly pion exchange we
should replace these \(3 \times 3\) matrices by isospin matrices
\(\vec{\tau} _{1,2} \). For a pair of nonstrange quarks the
difference between the two is small (see below). As noted already
in the previous section, if one is interested only in the mixing
angles (and not in fitting mass spectra) the value of the constant
B is immaterial, as we shall see. The main difference between Eq.\
(\ref{1}) and Eq.\ (\ref{2}) is the extra factor of
\(\vec{\lambda_{1}} \cdot \vec{\lambda_{2}}\) in \(H_{{\rm OPE}}\)
and the coefficient of the contact term relative to the tensor
term \((8 \pi/3\) versus \(-4 \pi /3\), a factor of minus one half
to go from OGE to OPE). It is interesting to note that the
coefficient of \((-4 \pi /3\)) in the pion exchange case is the
same as for the interaction of two electric dipoles
\cite{jackson}. For finite mass pions there are corrections to
 Eq.\ (\ref{2}) which will be discussed elsewhere (J. Chizma, to be published). \\

\subsection{\label{npe}Negative Parity Eigenstates}

The low mass negative parity baryons are assigned to a 70-plet of
\(SU(6)\) which means that the spatial wavefunctions have mixed
permutational symmetry (see \cite{karl, isgur, close}). In the
notation we use \cite{karl}, the spatial wavefunction \(\psi\) has
two components \(\psi^{\lambda}\) and \(\psi^{\rho} \) which
transform under permutations of the three quarks as a two
dimensional irreducible representation. The notation is explained
for example in \cite{close}. The total wavefunction \(\Psi\) is a
sum of products of spatial \(\psi\), spin \(\chi\) and flavor
\(\phi\) wavefunctions. For spin 3/2, the spin wavefunction is
totally symmetric under permutations while for spin 1/2 there are
again two states of mixed symmetry \(\chi^{\lambda}\) and
\(\chi^{\rho}\). The flavor wavefunctions for I=1/2, also have
mixed symmetry. Ignoring the color wavefunction (which is
antisymmetric) the total wavefunction is totally symmetric under
all permutations, and has the following forms:
\begin{subequations}\label{3}
\begin{eqnarray}
S=3/2: &    \Psi (^{4} P) & = \frac{1}{\sqrt{2}} \chi^{s} \lbrace
\psi^{\lambda} \phi^{\lambda} + \psi^{\rho} \phi^{\rho} \rbrace,
\\ \nonumber
  \\ S=1/2: &    \Psi(^{2} P) & = \frac{1}{2} \lbrace
\chi^{\lambda} \psi^{\rho} \phi^{\rho} + \chi^{\rho}
\psi^{\lambda} \phi^{\rho} + \chi^{\rho} \psi^{\rho}
\phi^{\lambda} - \chi^{\lambda} \psi^{\lambda} \phi^{\lambda}
\rbrace.
\end{eqnarray}
\end{subequations}

\section{\label{comp}Computations}
The spin angular momentum S=1/2, 3/2 has to be coupled with the
orbital angular momentum L=1 to give the total angular momentum
J=L+S. As a result there are two states each at J=1/2 and J=3/2,
namely spin doublet and spin quartet: \(^{2}P_{1/2}, ^{4}P_{1/2}\)
and \(^{2}P_{3/2}, ^{4}P_{3/2}\).  The physical eigenstates are
linear combinations of these two states, and can be obtained by
diagonalizing the Hamiltonian (\(H_{{\rm OGE}}\) or \(H_{{\rm
OPE}}\)) in this space of states. For example, the \(J^{{\rm P}} =
3/2^{-}\) states are eigenstates of the matrix:

\begin{equation}
\left(\begin{array}{rr} \langle ^{4} P_{3/2} \vert H \vert
^{4}P_{3/2} \rangle & \langle
^{4} P_{3/2} \vert H \vert ^{2}P_{3/2} \rangle  \\
\langle ^{2} P_{3/2} \vert H \vert ^{4}P_{3/2} \rangle & \langle
^{2} P_{3/2} \vert H \vert ^{2}P_{3/2} \rangle
\end{array}\right)_{,}
\end{equation}\\
where H is either \(H_{{\rm OGE}}\) or \(H_{{\rm OPE}}\), which
are given in Eq.\ (\ref{1}) or Eq.\ (\ref{2}) for a single pair of
quarks. The total Hamiltonian sums over all three quark pairs, and
since the wavefunctions in Eq.\ (\ref{3}) are symmetric under all
permutations, we can pick a single pair of quarks \(H^{(12)}\) and
multiply the result by three. The computation of these matrix
elements are simple but a little tedious. We illustrate the
computation for the case of \(J^{{\rm P}} = 3/2^{-}\) given above,
successively for both gluon and pion exchange. Before we start, we
comment that the doublet-doublet matrix elements only receive
contributions from the contact terms, while the doublet-quartet
matrix elements only come from tensor terms. The quartet-quartet
matrix element receives contributions from both tensor and contact
terms. Thus the relative size of contact and tensor terms come
into play. We find

\begin{equation}\label{4}
\langle ^{4} P_{3/2} \vert H_{{\rm OPE}} \vert ^{4} P_{3/2}
\rangle = (\frac{3}{2}) \lbrace \langle \chi^{s} \psi^{\lambda}
\vert H^{12} \vert \chi^{s} \psi^{\lambda} \rangle \langle
\phi^{\lambda} \vert \vec{\lambda_{1}} \cdot \vec{\lambda_{2}}
\vert \phi^{\lambda} \rangle +  \langle \chi^{s} \psi^{\rho} \vert
H^{12} \vert \chi^{s} \psi^{\rho} \rangle \langle \phi^{\rho}
\vert \vec{\lambda_{1}} \cdot \vec{\lambda_{2}} \vert \phi^{\rho}
\rangle \rbrace,
\end{equation}
where the leading factor of 3/2 comes from the number of pairs and
the normalization in Eq.\ (\ref{3}). We further take note that

\begin{eqnarray}\label{5}
\langle \phi^{\lambda} \vert \vec{\lambda_{1}} \cdot
\vec{\lambda_{2}} \vert \phi^{\lambda} \rangle = 4/3 & {\rm and} &
\langle \phi^{\rho} \vert \vec{\lambda_{1}} \cdot
\vec{\lambda_{2}} \vert \phi^{\rho} \rangle = -8/3.
\end{eqnarray}
It is amusing that these flavor matrix elements have coefficients
similar to the contact interaction in pion or gluon exchange, but
this is a simple numerical coincidence. We also assume harmonic
oscillator spatial wavefunctions \(\psi^{\rho, \lambda}_{1M}\) in
common with \cite{karl, glozman}. Then there is a further
simplification: the contact term which contains a delta function
\(\delta^{3} (\vec{\rho})\) vanishes in the state \(\psi^{\rho} \)
and only receives a contribution in the state \(\psi^{\lambda}\).
The tensor term only survives in \(\psi^{\rho}\), which has unit
orbital angular momentum \(l_{\rho} =1\). As a result we can
write:

\begin{equation}\label{6}
H_{{\rm OGE}}(^{4}P) = (3/2) \lbrace \langle \chi^{s}
\psi^{\lambda} \vert H^{12} _{{\rm contact}} \vert \chi^{s}
\psi^{\lambda} \rangle + \langle \chi^{s} \psi^{\rho} \vert H^{12}
_{{\rm tensor}} \vert \chi^{s} \psi^{\rho} \rangle,
\end{equation}
where for one gluon exchange

\begin{eqnarray}\label{7}
\langle \chi^{s} \psi^{\lambda} \vert H^{12} _{{\rm contact}}
\vert \chi^{s} \psi^{\lambda} \rangle & = & A(8 \pi /3) \langle
\chi^{s} \vert \vec{S_{1}} \cdot \vec{S_{2}} \vert \chi^{s}
\rangle \langle \psi^{\lambda} \vert \delta ^{3} (\vec{\rho})
\vert \psi ^{\lambda} \rangle,   \\ \nonumber & = & (2/3)A
\alpha^{3} \pi^{-1/2}.
\end{eqnarray}
Here \(\alpha\) is an oscillator parameter; the corresponding
tensor term is

\begin{equation}\label{8}
\langle \chi^{s} \psi^{\rho} \vert H^{12} _{{\rm tensor}} \vert
\chi^{s} \psi^{\rho} \rangle = (8/15)A \alpha^{3} \pi^{-1/2}.
\end{equation}
Inserting Eqs.\ (\ref{7}), (\ref{8}) into Eq.\ (\ref{6}) we
obtain, (in agreement with \cite{karl})

\begin{eqnarray}
\langle ^{4} P_{3/2} \vert H_{{\rm OGE}} \vert ^{4} P_{3/2}
\rangle & = & (3/2) \lbrace (2/3) +(8/15) \rbrace,   \\ \nonumber
& = & (9/5) \: \: ({\rm in \: units} \: A \alpha^{3} \pi^{-1/2}).
\end{eqnarray}
Similarly for pion exchange one obtains

\begin{eqnarray}
\langle ^{4} P_{3/2} \vert H_{{\rm OPE}} \vert ^{4} P_{3/2}
\rangle & = & (3/2) \lbrace (-1/2)(2/3)(4/3) + (1)(8/15)(-8/3)
\rbrace,
  \\ \nonumber & = & (-14/5) \: \: ({\rm in \: units} \: B \alpha^{3}
\pi^{-1/2}).
\end{eqnarray}
Where the factor of (-1/2) is the change in contact term from Eq.\
(\ref{1}) to Eq.\ (\ref{2}). There are only two more matrix
elements (for each of pion and gluon exchange), and they are
\begin{subequations}
\begin{eqnarray}
\langle ^{2} P_{3/2} \vert H_{{\rm OGE}} \vert ^{4} P_{3/2} \rangle & = & (10)^{-1/2} A \alpha^{3} \pi^{-1/2},   \\
\langle ^{2} P_{3/2} \vert H_{{\rm OGE}} \vert ^{2} P_{3/2} \rangle & = & -A \alpha^{3} \pi^{-1/2},   \\
\langle ^{2} P_{3/2} \vert H_{{\rm OPE}} \vert ^{4} P_{3/2} \rangle & = & (-8/3)(10)^{-1/2} B \alpha^{3} \pi^{-1/2},   \\
\langle ^{2} P_{3/2} \vert H_{{\rm OPE}} \vert ^{2} P_{3/2}
\rangle & = & (-7/3) B \alpha^{3} \pi^{-1/2}.
\end{eqnarray}
\end{subequations}
With these matrix elements we find for OGE, the Hamiltonian for
J=3/2 to have the form

\begin{equation}
\left(\begin{array}{lr}
9/5 & 10^{-1/2} \\
10^{-1/2} & -1 \\
\end{array}\right)
\left( \begin{array}{cc} ^{4}P_{3/2} \\ ^{2}P_{3/2}
\end{array} \right)_{,}
\end{equation} \\
where we have omitted the common units \(A \alpha^{3}
\pi^{-1/2}\). We now find the mixing \(\sin \theta_{{\rm d}}
\simeq (10^{-1/2})/(14/5) = 0.11\) corresponding to a mixing angle
of \(\theta_{{\rm d}} = 6.3 ^{\circ}\), in agreement with
\cite{karl, isgur}. The definition we follow has the lowest energy
state (``\(E_{{\rm low}} = -1.035\)'') with composition: \(\vert
E_{{\rm low}} \rangle = -\sin \theta_{{\rm d}} \: \vert ^{4}
P_{3/2} \rangle + \cos \theta_{{\rm d}} \: \vert ^{2} P_{3/2}
\rangle \). This means that the lowest eigenstate of the matrix
above is: \(\vert J^{{\rm P}} = 3/2^{-};{\rm OGE} \rangle = -0.110
\: \vert ^{4} P_{3/2} \rangle + 0.994 \: \vert ^{2} P_{3/2}
\rangle\). We emphasize that this mixing is the same for all
possible values of the constant \(A \alpha^{3} \pi^{-1/2}\),
whether they fit the masses or not. Similarly for \(H_{{\rm
OPE}}\) we have to diagonalize the matrix:

\begin{equation}
\left(\begin{array}{lr}
-14/5 & -(8/3)10^{-1/2} \\
-(8/3)10^{-1/2} & -7/3 \\
\end{array}\right)
\left( \begin{array}{cc} ^{4}P_{3/2} \\ ^{2}P_{3/2}
\end{array} \right)_{,}
\end{equation} \\
With this matrix, the mixing angle \(\theta_{{\rm d}}\) is found
to be \(\theta_{{\rm d}} = -52.7 ^{\circ}\). This means that the
lowest eigenstate of \(H_{{\rm OPE}}\) has the composition:
\(\vert J^{{\rm P}} = 3/2^{-};{\rm OPE} \rangle = 0.796 \: \vert
^{4} P_{3/2} \rangle + 0.606 \: \vert ^{2} P_{3/2} \rangle\). This
is very different from the composition of the state with OGE
coupling, given above. Whereas with OPE coupling the lowest
\(3/2^{-}\) state is about 63\% spin-quartet, with OGE it is about
1\% spin quartet. The decay data favors a 1\% contamination
\cite{hey}. Furthermore, \cite{nathan,glozman} quote a mixing
angle \(\theta_{\rm d} = \pm 8 ^{\circ}\) for OPE which differs
substantially from \(-53^{\circ}\). Note that had we used a
coupling \(\vec{\tau_{1}} \cdot \vec{\tau_{2}} \) instead of
\(\vec{\lambda_{1}} \cdot \vec{\lambda_{2}} \) the OPE composition
would change slightly to \(0.78 \: \vert ^{4} P_{3/2} \rangle +
0.63 \: \vert ^{2} P_{3/2} \rangle \).

We give now briefly the corresponding numbers in the \(J^{\rm P}
=1/2^{-}\) sector, referring to the lowest energy states in units
of \(A \alpha^{3} \pi^{-1/2}\) (for OGE) and \(B \alpha^{3}
\pi^{-1/2}\) (for OPE):

\begin{eqnarray}
{\rm OGE}:\vert {E = -1.62} \rangle & = & 0.526 \: \vert ^{4}
P_{1/2} \rangle + 0.85 \: \vert ^{2} P_{1/2} \rangle ; \:
\theta_{\rm s} = -32^{\circ},   \\\nonumber  \\  {\rm OPE}: \vert
{E = -3.60} \rangle & = & -0.43 \: \vert ^{4} P_{1/2} \rangle +
0.903 \: \vert
^{2} P_{1/2} \rangle ; \: \theta_{\rm s} = +25.5^{\circ}.   \\
\nonumber
\end{eqnarray}
References \cite{glozman} and \cite{nathan} quote a mixing angle
\(\theta_{\rm s} = \pm 13^{\circ}\). The data \cite{hey} supports
a composition close to OGE and a mixing angle of \(-32^{\circ}\).
For reference we quote the OPE Hamiltonian in the \(1/2^{-}\)
sector in matrix form, (in units \( B \alpha^{3} \pi^{-1/2}\))

\begin{equation}
\left(\begin{array}{lr}
 2 &  8/3 \\
8/3 & -7/3 \\
\end{array}\right)
\left( \begin{array}{cc} ^{4}P_{1/2} \\ ^{2}P_{1/2}
\end{array} \right)_{.}
\end{equation} \\
Note that for OPE coupling the lowest lying state is predominantly
(81\%) spin doublet, while for OGE coupling the ground state is
72\% spin doublet.

\section{\label{sad}Summary and Discussion}

In addition to the spin-spin couplings of Eqn's (\ref{1}) or
(\ref{2}) discussed above, spin in the negative parity baryons
also couple to the orbital angular momentum. This is the
spin-orbit coupling ($\vec{L} \cdot \vec{S} $). It is an empirical
observation that this coupling is rather weak in negative parity
nucleons, and as a result it has been neglected in some of the
literature \cite{karl, isgur,glozman}. There is a great deal of
discussion about the physical origin of this effect
\cite{nathan,gloz}. If some spin orbit coupling is included it
will contribute to the diagonal matrix elements of the two by two
matrices which we diagonalize, and can shift the mixing angles.
The inclusion of this effect will however negate a parameter free
determination of the mixing angles. That is, one must use the
spectroscopic mass data in order to find the relative strengths of
the hyperfine interaction (Eqn's \ref{1},\ref{2}) and the spin
orbit interaction. This was done in a very preliminary manner, and
we find the changes to the mixing angles to be small - less than
the experimental error of $10 ^{\circ}$. The spectroscopic data
utilized was the nucleon, delta ($P_{33}$ resonance) and the
$D_{13}$ (low), $D_{15}$ mass splitting. With these splittings, we
find that the OGE mixing angle changes from $-32 ^{\circ}$ to $-36
^{\circ}$, while for OPE the angle changes from $25.5 ^{\circ}$ to
$27.5 ^{\circ}$ (both for the $J^{P} = 1/2 ^{-}$ sector). For more
complete results, one should attempt a reasonable fit to all
states in the multiplet and this has not yet been done.

We summarize in Table \ref{table1} the results quoted in section
\ref{comp}. We emphasize again that these results are independent
of spectral fits to the masses of these states. The results depend
only on the couplings and the wavefunctions assumed. The strength
of the coupling (either A or B here) factorizes from the mixing
matrices and the same mixing is obtained regardless of this
coupling strength. The wavefunctions were assumed to be harmonic
oscillator - appropriate for these states since they are the
ground state in the negative parity sector. Moreover this
assumption is common to \cite{karl, isgur} and \cite{glozman}. We
further assumed in the couplings for OPE that the pion mass is
zero; results for non-zero pion masses will be given elsewhere.
They do not change significantly the numbers given in Table
\ref{table1}.

Table \ref{table1} shows a substantial change from the values for
OPE in the literature \cite{glozman,nathan}; this change is
relevant since the error of the ``experimental'' value is of the
order of \(10^{\circ}\) \cite{hey}, and the preference for the OGE
solution is now unambiguous. It has been argued \cite{gloz} that
the addition of vector meson exchange to pseudoscalar exchange
will remedy this problem. That may indeed be the case, but one
should recall that the primary controversy is whether the quark
coupling in baryons is OGE or OPE, and the data answers this
question unequivocally. One may just as well argue that atomic
hyperfine interactions - which has the same form as Eq.\ (\ref{1})
- is really due to the superposition of a pseudoscalar and massive
vector field, rather than a massless gauge field. Similar mixings
for OPE have also been obtained elsewhere \cite{arima}, however
the emphasis on the independence from spectral data is missing.

Finally, there have been comments on the issue of color versus
flavor exchange. In particular, \cite{collins} fits the mass
spectrum in the L=1 sector in a rather ingenious way, using only
permutation symmetry and SU(6), with a number of free parameters,
essentially reduced matrix elements. But by treating the matrix
elements corresponding to the contact and tensor terms as
independent parameters, one sidesteps the controversy between
vector exchange \cite{georgi,karl,isgur} and pseudoscalar exchange
\cite{glozman}. In addition, as noted, if one has well defined
Hamiltonians and wavefunctions these mixings are independent of
mass fits. Although not in these precise words, similar
conclusions are stated by the authors of \cite{collins}.

\begin{table}
\caption{\label{table1} \\ Summary of Results}
\begin{ruledtabular}
\begin{tabular}{llccr}

& Coupling & Reference & Mixing Angle & $\%^{4}P_{j}$ \\ \hline \\
$J^{\rm P}$ = $3/2^{-}$ & OPE & 6, 7 & $\pm 8^{\circ}$ & $2 \%$ \\
& OPE & this ref. & $-52.7 ^{\circ}$ & $63\%$ \\
& OGE & 3, 4 \& this ref. & $+6 ^{\circ}$ & $1 \%$ \\
& EXP. & 11 & $+10 ^{\circ}$ & $3 \%$ \\ \\
$J^{\rm P}$ = $1/2^{-}$ & OPE & 6, 7 & $\pm 13^{\circ}$ & $5 \%$ \\
& OPE & this ref. & $+25.5 ^{\circ}$ & $19 \%$ \\
& OGE & 3, 4 \& this ref. & $-32 ^{\circ}$ & $28 \%$ \\
& EXP. & 11 & $-32 ^{\circ}$ & $28 \%$ \\

\end{tabular}
\end{ruledtabular}
\end{table}

\begin{acknowledgments}
We would like to thank Prof. Victor Novikov for comments on the
manuscript. We also thank Masanori Morishita and Masaki Arima for
pointing out a sign error in the first version of this manuscript.
\end{acknowledgments}


\begin{thebibliography}{99}

\bibitem{georgi} A. DeRujula, H. Georgi and S.L. Glashow,
Phys. Rev. D \textbf{12}, 147 (1975).

\bibitem{gromes} D. Gromes and I.O. Stamatescu,
Nucl. Phys. B \textbf{112}, 213 (1976).

\bibitem{karl} Nathan Isgur and Gabriel Karl,
Physics Letters B \textbf{73}, 109 (1977).

\bibitem{isgur} Nathan Isgur and Gabriel Karl,
Phys. Rev. D \textbf{18}, 4187 (1978).

\bibitem{brown} see eg., G.E. Brown, Mannque Rho,
Phys. Lett. B \textbf{82}, 177 (1979); \\
A.W. Thomas, S. Theberge, G.A. Miller, Phys. Rev. D\textbf{24},
216 (1981).

\bibitem{glozman} L.Ya Glozman and D.O. Riska,
Physics Reports \textbf{268}, 263 (1996).


\bibitem{nathan} Nathan Isgur,
Phys. Rev. D \textbf{62} , 054026 (2000).

\bibitem{gloz} For a response to Ref.\ \cite{nathan}, see: L.Ya Glozman,
 nucl-th/9909021.

\bibitem{jackson} J.D. Jackson,
\emph{Classical Electrodynamics}, Second Edition (John Wiley and
Sons, 1975), pages 141 and 187.

\bibitem{close} F.E. Close,
\emph{Introduction to Quarks and Partons} (Academic Press, 1979);
 G. Karl, \emph{Elements of Baryon Spectroscopy} in \emph{Progress
in Particle and Nuclear Physics}, 1985, edited by A. Faessler,
(Pergamon Press), pages 243-253.

\bibitem{hey} A.G.H. Hey, P.J. Litchfield and R.J. Cashmore,
Nucl. Phys. B \textbf{95}, 516 (1975).

\bibitem{arima} M. Morishita and M. Arima,
Phys. Rev. C \textbf{65}, 045209 (2002); see also M. Arima, K
Shimizu and K. Yazaki, Nucl. Phys. A \textbf{543}, 613 (1992).

\bibitem{collins} H. Collins and H. Georgi,
Phys. Rev. D \textbf{59}, 094010 (1999).

\end{thebibliography}
\end{document}